\documentclass[aps,prl,twocolumn,epsfig,groupedaddress]{revtex4}

\usepackage{epsfig}
\begin{document}

\title{Energy Dependence of the Cronin Effect from Non-Linear QCD 
Evolution\footnote{We dedicate this work to the memory of Ian Kogan.}}

\author{Javier L. Albacete$^{1,2}$,
N\'estor Armesto$^{2}$, Alex Kovner$^{2,3}$,
Carlos A. Salgado$^{2}$ and Urs Achim Wiedemann$^{2}$}
\address{
$^1$ Departamento de F\'{\i}sica, Universidad de C\'ordoba,
14071 C\'ordoba, Spain\\
$^2$ Theory Division, CERN, CH-1211 Geneva 23, Switzerland\\
$^3$ Department of Mathematics and Statistics, University of
Plymouth, Drake Circus,
Plymouth PL4 8AA, UK}
\date{\today}

\begin{abstract}
The non-linear evolution of dense partonic systems has been suggested
as one of the novel physics mechanisms relevant to 
the dynamics of hadron--nucleus and nucleus--nucleus collisions at
collider energies. Here we study to what extent the description
of Cronin enhancement in the framework of this non-linear evolution 
is consistent with the recent observation in $\sqrt{s} = 200$ GeV 
d--Au collisions at the Relativistic Heavy Ion Collider.  
We solve the Balitsky-Kovchegov (BK) evolution equation 
numerically for several initial conditions encoding Cronin enhancement. 
We find that the properly normalized nuclear gluon distribution is 
suppressed at all momenta relative to that 
of a single nucleon. Calculating the resulting spectrum of produced
gluons in p--A and A--A collisions, we establish that the nonlinear 
QCD evolution is unable to generate a Cronin type enhancement, 
and that it quickly erases any such enhancement which may be present at 
lower energies.  
\end{abstract}
\maketitle
 \vskip 0.3cm

%%%%%%%%%%%%%%%%%%%%%%%%%%%%%%%%%%%%%%%%%%%%%%%%%%%%%%%%%%%%%%%%%%%%%%
The observation that the ratio of particle yields in p--A and A--A, scaled 
by the number of collisions, exceeds unity in an intermediate transverse
momentum range of a few GeV, is commonly referred to as the Cronin effect. 
This was first seen at lower fixed target energies~\cite{Cronin:zm} and 
was recently confirmed in $\sqrt{s} = 200$ GeV d--Au collisions 
at RHIC~\cite{dA}. The current interest focuses mainly on comparing
this Cronin {\it enhancement} in d--Au to the relative {\it suppression} of 
produced hadrons in Au--Au collisions at the same center of mass energy
and in the same transverse momentum range~\cite{gold}. The opposite
trend of the two effects and their centrality dependence
suggests that d--Au data may serve as an efficient benchmark measurement
to distinguish between the two different physical mechanisms 
suggested for the relative suppression of hadron spectra in Au--Au collisions:
initial state parton saturation\cite{klm} and final state jet 
quenching~\cite{Wang:2003aw}. 

The physics of dense partonic systems and their non-linear perturbative
evolution to higher energy has motivated several attempts
at understanding bulk properties of ultra-relativistic heavy ion
collisions such as the multiplicity, rapidity distribution and
centrality dependence of particle production
\cite{kln,klm}. In particular it has been suggested that 
saturation effects can account for the suppression of the high-$p_T$ 
hadronic spectra in Au--Au collisions at RHIC. On the other hand, it is
known that saturation models based on multiple scattering 
(the so called Glauber-Mueller~\cite{glaubermuller} or 
McLerran-Venugopalan~\cite{mv} models) exhibit Cronin 
enhancement in p--A~\cite{Gelis:2002nn,bkw,kkt}
and A--A~\cite{bkw,Jalilian-Marian:2003mf}. 
In these models, the saturation 
of low $p_T$ gluons is the result of a redistribution of 
gluons in transverse phase space~\cite{kovchegov,jkmw} which does not 
change the total number of gluons, thus resulting in a compensating 
enhancement at momenta just above the saturation momentum $Q_s$. What 
is not fully understood is i) whether such Cronin enhancement encoded 
in the initial condition of a nuclear wave function persists in the 
non-linear perturbative QCD evolution to higher energy and ii) whether 
such Cronin enhancement can be generated by the non-linear 
evolution itself. This paper goes beyond earlier 
discussions~\cite{bkw,kkt,Jalilian-Marian:2003mf} by
providing the first complete (numerical) 
answer to these questions. We do not address other approaches
to Cronin enhancement~\cite{othercronins}.

We start from the Balitsky-Kovchegov (BK) evolution 
equation~\cite{balitsky,k}, which describes the evolution of the forward
scattering amplitude $N({\bf r},y)$ of a QCD dipole of transverse size 
$|{\bf r}|$ with rapidity $Y$ and $y = (\alpha_s\, N_c/\pi)\, Y$,
% . For a large We consider the translational and rotational invariant 
% limit, in which the BK-equation does not depend on impact parameter and 
% the direction of the vector ${\bf r}$ 
%
\begin{eqnarray}
  &&{dN(|{\bf r}|,y)\over dy} = {1\over 2\pi}\int d^2{\bf z}\,
     {({\bf r}-{\bf z})\cdot{\bf z}\over ({\bf r}-{\bf z})^2\, {\bf z}^2}
  \label{bk} \\
 && \times [N(|{\bf r}-{\bf z}|)+N(|{\bf z}|)- N(|{\bf r}|)-
      N(|{\bf r}-{\bf z}|)N(|{\bf z}|)]\, .
  \nonumber
\end{eqnarray}
The unintegrated gluon distribution is related to the inclusive gluon 
distribution $\phi(k) \propto {d(xG(x,k^2))\over d^2k\, d^2b}$
and is given in terms of 
the dipole amplitude 
\begin{equation}
   \phi(k) 
   = \int {d^2r\over 2\pi\, r^2}\exp\{i\, {\bf r}\cdot {\bf k}\}N(r)\, .
   \label{phi}
\end{equation}
In the following, we also use the modified gluon distribution
\begin{equation}
  h(k)=k^2\, \nabla^2_k\, \phi(k)\, .
  \label{hhh}
\end{equation}
The two definitions coincide for the leading order perturbative 
distribution $\phi(k)\propto {1\over k^2}$, but are different in 
general, and especially at low momenta. 

Using the second order Runge-Kutta algorithm~\cite{Braun:2000wr}, we solve 
the BK equation (\ref{bk}) numerically with 8000 equally spaced 
intervals in $\ln k$-space between $-15$ and $35$ and a step
$\Delta\, y = 0.0025$. The accuracy of this algorithm is better 
than 2 \% in the entire range of $k$ discussed below. We evolve
two initial conditions given by the McLerran-Venugopalan \cite{mv} (MV) 
and Golec-Biernat--W\"usthoff \cite{gbw} (GBW) model respectively:
\begin{eqnarray}
  N^{Q_s}_{MV}&=& 1-
  \exp\left[-\frac{Q_s^2r^2}{4}
        \ln\left({ 1\over r^2\Lambda^2_{QCD}}+e\right)\right]\, ,
  \\
  N^{Q_s}_{GBW}&=&1-\exp\left[-\frac{Q_s^2r^2}{4}\right]\, ,
\end{eqnarray}
where $\Lambda_{\rm QCD} = 0.2~{\rm GeV}$.
For momenta $k \ge O(1 ~{\rm GeV})$, 
the sensitivity on the infrared cut off $e$ is negligible. The
amplitudes $N_{MV}$ and $N_{GBW}$ are similar for momenta of order 
$Q_s$, but differ strongly in their high $k$ behavior; $\phi_{GBW}(k)$ 
decays exponentially while $\phi_{MV}$ has a power-like tail $1/k^2$.

\begin{figure}[h]\epsfxsize=8.7cm
\centerline{\epsfbox{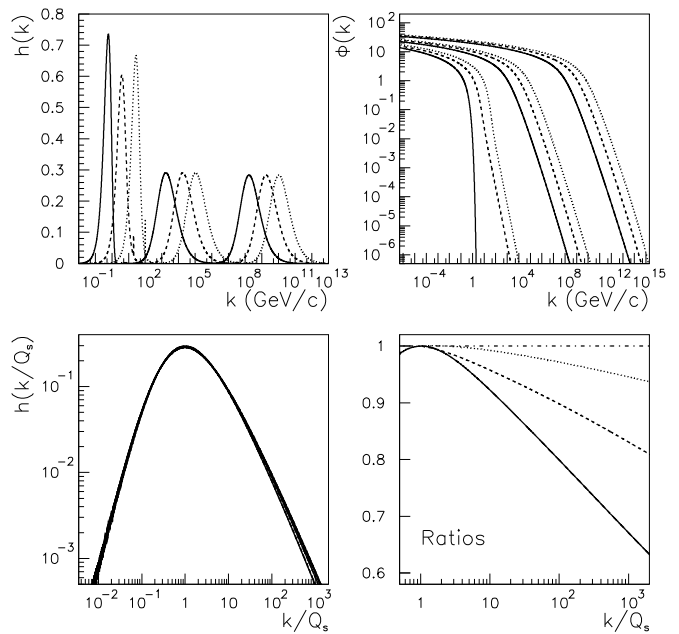}}
%\vspace{0.5cm}
\caption{ Solutions of the BK equation.
{\it Upper-left}: $h(k)$ evolved (left to right) from $y=0$ to
5 and 10 for different initial conditions: GBW with $Q_s^2=0.36$ GeV$^2$ 
(solid lines), MV with $Q_s^2=4$ GeV$^2$ (dashed lines) and MV with 
$Q_s^2=100$ GeV$^2$ (dotted lines).
{\it Upper-right}: The same as upper left for $\phi(k)$. {\it
Lower-left}: the scaled function
$h(\rho)$ versus $\rho = k/Q_s$ for $y=4,6,8,10$ and the same
initial conditions and conventions (lines cannot be distinguished). {\it
Lower-right}: Ratio of $h(y,\rho)/h(y,\rho=1)$ over
$h(y=10,\rho)/h(y=10,\rho=1)$ for $y=4$ (solid line), 6 (dashed line), 8
(dotted line)  and 10 (dashed-dotted line), and initial condition MV with
$Q_s^2=4$ GeV$^2$.
}\label{fig1}
\end{figure}

Fig.\ref{fig1} shows the evolution of $h(k,y)$ and $\phi(k,y)$ for 
different initial conditions.
The solutions for $h(k,y)$ quickly approach a universal soliton-like 
shape and do not change further except uniformly moving in $k$ 
on the logarithmic plot.
The position of the maximum
is the evolved value of the saturation momentum $Q_s(y)$.
The solutions for different initial conditions and different 
rapidities scale as a function of the scaling variable $\rho=k/Q_s(y)$. 
The shape of the initial condition affects only the value of the 
saturation momentum $Q_s(y)$, but not the shape of the 
evolved function $h(\rho,y)$. 
The $y$-dependence of $h(\rho,y)$ is very weak: the function evolves fast  
towards a scaling form $h(\rho)$. As the rapidity changes between
$y=4$ and $y=10$, the ratio $h(\rho,y_1)/h(\rho,y_2)$ varies by at most 
$40\%$ over three orders of magnitude of the scaling variable $\rho$.
Similar behavior was found for $\phi$ (results not shown). This is 
consistent with previous numerical works~\cite{ab,gbs}.
%
%%%%%%%%%%%%%%%%%%%%%%%%%%%%%%%%%%%%%%%%%%%%%%%%%%%%%%%%%%%%%%%%%%%%
%%%%%%%%%%%%%%%%%%%%%%%%%%%%%%%%%%%%%%%%%%%%%%%%%%%%%%%%%%%%%%%%%%%
%

To get a quantitative idea of the behavior of the scaling solution, 
we fitted the numerical solution of $\phi(\rho)$ to two analytical
expressions: $s_1(\rho)=a\rho^{2(1-\lambda)}$
and $s_2(\rho)=a\ln (b\rho)\rho^{2(1-\lambda)}$ for $\rho > 5$.
The functional form $s_1$ with $\lambda=0.37$ and $\ln Q_s \propto y$
describes the scaling
behavior of solutions of the linear BFKL equation~\cite{i^2m}. 
It was argued in Ref.~\cite{mt} that $s_2$ with the same value of 
$\lambda$ and $\ln Q_s \propto \frac{2\chi(\lambda)}{1-\lambda} y
- \frac{3}{2(1-\lambda)}\, \ln y$
accounts for the effects of nonlinearities in Eq. (\ref{bk}).
We find that $s_1$ does not give an acceptable fit to $\phi(\rho)$
in any extended range of $\rho$. For values of $\rho$ between $1$ and 
$10^3$ the value of $\lambda$ varies between 0.39 and 0.46. This is in 
contrast to the BFKL equation, where we find numerically that $s_1$ 
with $\lambda=0.37$ does indeed approximate the
solution over several orders of magnitude with very 
good accuracy (results not shown). On the other hand, for $5 < \rho < 1000$,
$s_2$ gives a good fit with $\lambda=0.32$. 
If, following \cite{mt} we fix $\lambda=0.37$, the fit 
is still good. 
%
%%%%%%%%%%%%%%%%%%%%%%%%%%%%%%%%%%%%%%%%%%%%%%%%%%%%%%%%%%%%%%%%%%%%
\begin{figure}[h]\epsfxsize=8.7cm
\vspace*{-0.6cm}
\centerline{\epsfbox{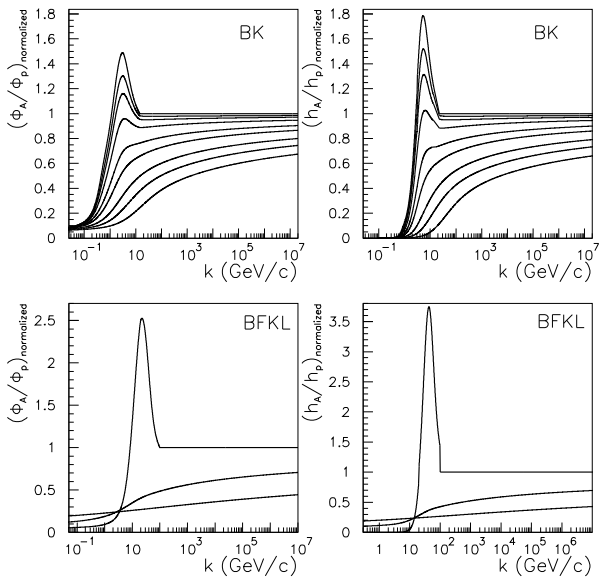}}
%\vspace{0.5cm}
\caption{ Ratio of distributions $\phi$ and $h$ in nucleus
and proton, normalized to 1 at $k \to \infty$.  {\it Upper plots}:  BK
evolution, with MV as initial condition with $Q^2_s=0.1$ GeV$^2$ for p and
2 GeV$^2$ for A. Lines from top to bottom correspond to $y=0$, 0.05,
0.1, 0.2, 0.4, 0.6, 1, 1.4 and 2. {\it Lower plots}: BFKL evolution, with
MV as initial condition with $Q_s^2=4$ GeV$^2$ for p and 100 GeV$^2$ for
A. Lines from top to bottom correspond to $y=0$, 1 and 4.
}\label{fig2}
\end{figure}
%%%%%%%%%%%%%%%%%%%%%%%%%%%%%%%%%%%%%%%%%%%%%%%%%%%%%%%%%%%%%%%%%%%

To study the effect of the evolution on the Cronin enhancement, we 
consider two initial conditions, $N_{MV}^{q}(r)$ and $N_{MV}^{Q}(r)$ 
with $q^2=0.1\, {\rm GeV}^2$ and $Q^2=2\, {\rm GeV}^2$. Since 
$q \sim \Lambda_{QCD}$ and $Q$ is of order of the estimated saturation 
momentum for a gold nucleus~\cite{kln}, this choice mimics the gluon 
distributions of a proton and of a nucleus respectively. 
At large transverse momenta the ratio of the corresponding Fourier 
transforms is given by the ratio of the saturation momenta,
\begin{equation}
  {h^{Q}(k,y=0)\over h^{q}(k,y=0)}={Q^2\over q^2} = A^{1/3}\, .
\end{equation}
This relation also holds for $\phi$.
As discussed in \cite{bkw,kkt}, these initial conditions exhibit 
Cronin enhancement, namely
${h^Q(k,y=0)\over A^{1/3}h^q(k,y=0)}> 1$ for $k\sim Q$. 
We solve the BK equation with these two initial conditions and 
construct the ratio $R(k,y)=h^Q(k,y)/A^{1/3}h^q(k,y)$
and the corresponding ratio for $\phi$, see Fig.~\ref{fig2}. 
The initial Cronin enhancement at rapidity $y=0$ is seen to be wiped 
out very quickly by the evolution. Within less than half a  
unit of rapidity $y$ the ratios 
show uniform suppression for all values of transverse 
momentum. The observed behaviour persists if different 
amounts of Cronin enhancement are included in the initial condition.

As seen in the lower panel of Fig.~\ref{fig2}, the Cronin enhancement
also disappears rapidly with rapidity when gluon distributions are 
evolved according to the linear BFKL equation. Qualitative differences
between the BFKL and BK dynamics are only visible at momenta $k< Q_s$, 
where saturation effects are important. For larger momenta $k$, the
ratios are very similar for 
linear and non-linear QCD evolution. We thus conclude that the wiping 
out of the initial enhancement is primarily driven by the linear BFKL 
dynamics which is contained in the BK equation as well. 

\begin{figure}[h]\epsfxsize=8.7cm
\vspace*{-0.6cm}
\centerline{\epsfbox{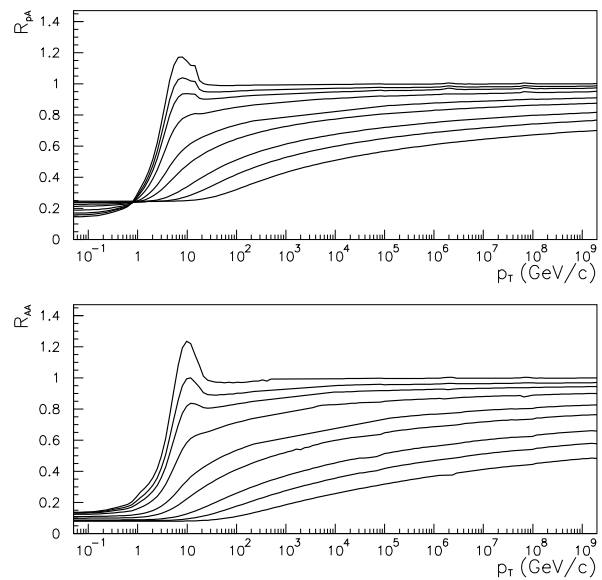}}
%\vspace{0.5cm}
\caption{Ratios $R_{pA}$ and $R_{AA}$ of gluon yields in p--A (upper plot) 
and A--A (lower plot) for BK evolution, with MV
as initial condition with $Q^2_s=0.1$ GeV$^2$ for p and 2 GeV$^2$ for A.
Lines from top to bottom correspond to $y=0$, 0.05, 0.1, 0.2, 0.4, 0.6, 1,
1.4 and 2.
}\label{fig3}
\end{figure}

For the evolved gluon distributions determined above, we have calculated
the yield of produced gluons in p--A and A--A 
collisions at central rapidity according to the factorized 
expressions \cite{glr}
\begin{eqnarray}
   {dN_{pA}\over dyd^2p\, d^2b}
   &\propto& {1\over p^2}
    \int d^2k\, h^q(y,k)\,h^Q (y,p-k)\, ,
   \label{pA}\\
   {dN_{AA}\over dy\, d^2p\, d^2b}
   &\propto&  {A^{2/3}\over p^2}
    \int d^2k\, h^Q(y,k)\, h^Q(y,p-k)\, .
   \label{AA}
\end{eqnarray}
From these spectra, we calculate the $p$- and $y$-dependent ratios
\begin{eqnarray}
   R_{pA}&=&{ {dN_{pA}\over dyd^2p\, d^2b}\over A^{1/3} 
     {dN_{pp}\over dyd^2p\, d^2b}}\, , \qquad
%     \label{rpa}\\
%
   R_{AA} = { {dN_{AA}\over dyd^2p\, d^2b}\over A^{4/3} 
            {dN_{pp}\over dyd^2p\, d^2b}}\, .
          \nonumber
%     \label{raa}\, .
\end{eqnarray}
As seen in Fig.~\ref{fig3}, the non-linear BK evolution quickly
wipes out any initial Cronin enhancement not only on the level
of single parton distribution functions but also on the level of 
particle spectra. We have checked that this behaviour is generic
by evolving different initial conditions corresponding to different 
initial amounts of enhancement. We note that in calculations of 
the gluon production in p--A in the eikonal
approximation \cite{kovchegov,km,braun}, the gluon distribution
$h$ rather than $\phi$ enters the right hand side of (\ref{pA}),
but no similar statement exists of nucleus-nucleus collisions.
We have checked that the results using $\phi$ are very close
to those shown in Fig.~\ref{fig3}. More generally,
the expressions (\ref{pA}) and (\ref{AA}) 
are based on rather strong approximations discussed in Ref.~\cite{bkw}.
However, our conclusion about the disappearance of Cronin
enhancement during QCD evolution is likely to persist in more refined
ways of calculating particle spectra, since it is rooted directly in
the rapidity dependence of gluon distributions.  
%
%%%%%%%%%%%%%%%%%%%%%%%%%%%%%%%%%%%%%%%%%%%%%%%%%%%%%%%%%%%%%%%%%%%%
%%%%%%%%%%%%%%%%%%%%%%%%%%%%%%%%%%%%%%%%%%%%%%%%%%%%%%%%%%%%%%%%%%%
%

We now comment on a recent formal argument~\cite{kkt} which 
- in contrast to our numerical findings - suggests that Cronin 
enhancement survives 
the non-linear evolution. It is based on the observation that at 
very short distances $r\rightarrow 0$, the dipole amplitude $N(r)$ 
is not affected by evolution. Thus, the integral of the gluon 
distribution function $\phi$ over the transverse momentum 
is expected to be rapidity independent,
\begin{equation}
  \int d^2k\, \phi(k)={1\over r^2}\, N(r)|_{r=0}\, .
\label{sum}
\end{equation}
One thus obtains the sum rule
\begin{equation}
\int d^2k\, \phi_A(k,y)=A^{1/3}\int d^2k\,\phi_p(k,y)
\label{sum1}
\end{equation}
valid for any rapidity, since it is satisfied by the initial condition 
$\phi_{MV}$. Since the nonlinear evolution leads to the depletion of the 
gluon distribution $\phi_A(k)$ relative to $A^{1/3}\phi_p(k)$ at low 
momenta, it must follow that in some range of momenta this effect is 
compensated by enhancement of $\phi_A$. Similar arguments can be made 
about the distribution $h$ and also about the gluon yield in p--A and 
A--A collisions.

However, this argument breaks down since the quantity defined in 
Eq. (\ref{sum}) is infinite.
As such Eq. (\ref{sum1}) relates only the 
divergent parts of the integrals which are dominated by ultraviolet, and
carries no information about possible behaviour at finite momentum.
To be more specific, we use the scaling property
$\phi(k,y)=\phi(k/Q_s(y))$ of the solution of the BK equation
established above. It is known that the ratio of the saturation 
momenta for any two solutions is preserved by the BK 
evolution \cite{ab,gbs,i^2m}. For our two solutions representing 
a nucleus and a nucleon, this implies ${Q^A_s(y)\over Q^p_s(y)}=A^{1/6}$.
We now rewrite the sum rule (\ref{sum1}) by regulating the divergent
integrals with a large but finite UV cutoff $aQ_s^A(y)$, 
\begin{eqnarray}
  &&\int_0^{a^2(Q_s^A)^2} d^2k \phi_A(k,y)=
  (Q_s^A)^2\int_0^{a^2} d^2\rho \phi(\rho)
  \label{nosumrule} \\
  && =A^{1/3}(Q_s^p)^2\int_0^{a^2} \hspace*{-0.3cm} d^2\rho \phi(\rho)=
  A^{1/3}\int_0^{a^2(Q_s^p)^2}  \hspace*{-0.3cm} d^2k \phi_p(k,y)\, .
  \nonumber
\end{eqnarray}
In the formal limit $a\to \infty$, we recover Eq. (\ref{sum1}). However,
since $Q_s^A \gg Q_s^p$, the regularized sum rule (\ref{nosumrule}) is easily 
satisfied even if the 
nuclear distribution is suppressed relative to that of a single nucleon 
uniformly at all momenta. Thus, the sum rule (\ref{sum1}) carries no 
information about either presence or absence of the Cronin enhancement.

In summary, we have found that the nonlinear QCD evolution to 
high energy is very 
efficient in erasing any Cronin type enhancement which may be present 
in the initial conditions. For realistic initial conditions this 
disappearance occurs within half a unit of rapidity. We note that in 
our units the evolution from $130$ GeV to $200$ GeV corresponds to 
$\delta y\simeq 0.1$ for $\alpha_s = 0.2$, and thus is not sufficient to 
completely eliminate an initial enhancement at central rapidity. For 
forward rapidity, $\delta y$ is greater.
The evolution to the LHC energy corresponds to 
$\delta y \sim 1$. Thus the BK evolution suggests the reduction
of the Cronin effect in d--Au for forward rapidities at RHIC and
it predicts the disappearance of the Cronin effect for p--A 
collisions at LHC.

We thank the CERN TH Division (JLA and AK) and the INT Seattle 
for its hospitality and the DOE for partial support (NA, AK, CAS and UAW).
JLA is supported by MECD of Spain, grant no. AP2001-3333, and 
CAS by a Marie Curie Fellowship no.
HPMF-CT-2000-01025 of the European Community TMR program.
Useful discussions with R.~Baier, E.~Iancu, D.~Kharzeev, 
Y.~Kovchegov, P.~Jacobs and L.~McLerran are 
gratefully acknowledged. We thank M.~A.~Braun for discussions and
a numerical cross-check.
%
%%%%%%%%%%%%%%%%%%%%%%%%%%%%%%%%%%%%%%%%%%%%%%%%%%%%%%%%%%%%%%%%%%%%%%%

\end{document}